\documentstyle[11pt,paspconf,psfig]{article}

\begin{document}

\title{Determination of the transverse velocity of Q2237+0305}

\author{J. Stuart Wyithe\altaffilmark{1}}
\affil{The University of Melbourne, Parkville, Vic, 3052, Australia} 

\altaffiltext{1}{Visiting Student Research Collaborator, Princeton University Observatory, Peyton Hall, Princeton, NJ 08544, USA}

\begin{abstract}
The largest systematic uncertainty present in the analysis of gravitationally microlensed quasar light curves is that of the galactic transverse velocity. We describe a method for determining the transverse velocity as well as its application to published monitoring data of Q2237+0305 (Irwin et al. 1989; Corrigan et al. 1991; $\O$stensen et al 1995). We find that while this data displays strong evidence for microlensing, it limits the transverse velocity to be $<500\,km\,sec^{-1}$ (for microlens masses of $0.5M_{\odot}$). In addition, by combining these results with a new method for modelling the microlensing contribution of stellar proper motions, we find that the masses of microlenses in Q2237+0305 are consistent with those of low mass stars.
\end{abstract}


\section{Introduction}
Q2237+0305 is a multiply imaged, gravitationally microlensed quasar. In addition to the difficulties presented by the relatively rare occurrence of high magnification events (HME), the analysis of published monitoring data of Q2237+0305 through observation of HME characteristics and rates is hampered by the sparse sampling rate. Previous light-curve peaks are under sampled and may have been missed altogether. However, comparatively speaking the monitoring data provides a good record of the microlensing rate described by the longer timescale low level fluctuations.

\section{Determination of effective galactic transverse velocity}

The observed microlensing rate is produced by the combination of microlens proper motions and a galactic transverse velocity. We define the effective transverse velocity as being that which in combination with a static microlensing model, produces a microlensing rate equal to that of the observed light curve (Wyithe, Webster \& Turner 1999b,c). The effective transverse velocity is therefore larger than the physical transverse velocity. 

\begin{figure}
\centerline{\psfig{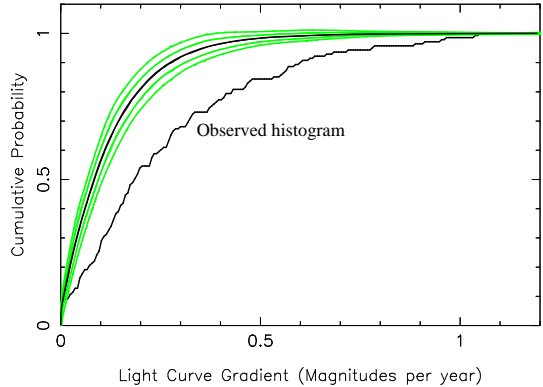}}
\caption{\label{noise}The cumulative histogram for the monitoring data (labelled), and histograms representing the mean (dark line) and $\pm1\sigma$, $\pm2\sigma$ levels (light lines) of fluctuation due to noise only.}
\end{figure}

  We assume that the macro models (eg. Schmidt, Webster \& Lewis 1998) for the lensing galaxy in the Q2237+0305 system correctly describe the shear and optical depth parameters at the position of each image. We assume that the source is small with respect to the microlens Einstein radius (eg Wambsganss Paczynski \& Schneider 1990), and find that (in this regime) limits on effective transverse velocity and microlens mass are approximately independent of the source size assumed. 

\subsection{The distribution of microlensed light-curve derivatives}

We are interested in the independent fluctuations in the 4 individual images of Q2237+0305. It is advantageous to look at the rates of change of the differences between image magnitudes because intrinsic fluctuation (time-delays are less than 1 day (eg. Schneider et al. 1988)) and the systematic component of error are removed.

The basic tool of our analysis is the distribution of difference light curve derivatives (Wyithe, Webster \& Turner 1999a,b,c) in which the transverse velocity is a scaling factor.
 The cumulative histogram has a smaller value if the transverse velocity is higher. Thus by considering the shape of the cumulative distribution of light curve derivatives, information on the transverse velocity can be obtained.

\subsection{Monitoring data}

 By combining the data of Kent \& Falco (1988), Schneider et al. (1988), Irwin et al. (1989), Corrigan et al. (1991) and $\O$stensen et al. (1996), a light curve is produced for each of the four images. The most complete light curves are in R-band. There are a total of 61 published data points for each image taken between 1985, and 1996. To minimise the noise in the data, we averaged adjacent observations leaving 26 points. We simulate errors in the model light curves according to a Gaussian distribution. $\Delta m=0.02$ is taken as the $2\sigma$ and $1\sigma$ level in images A/B and C/D respectively. The level of noise in the sample is estimated by calculating the fluctuations that result from the application of the observational uncertainty and sampling rate to a flat light curve.
 Figure \ref{noise} shows the cumulative histogram of difference light-curve derivatives for Q2237+0305 together with cumulative histograms for the mean, $\pm1\sigma$ and $\pm 2\sigma$ cumulative histograms of derivatives due to observational error only. The difference between these histograms demonstrates a statistically significant level of microlensing.

\subsection{Limits on effective transverse velocity}

\begin{figure}
\centerline{\psfig{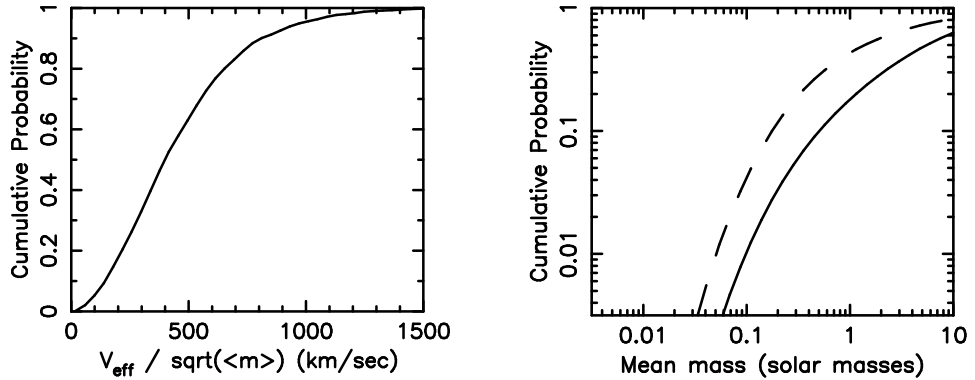}}
\caption{\label{velmass}Left: $P_{v}(V_{gal}<V_{eff})$ and Right: $U\left(\langle m\rangle<M\right)$ under assumption of flat prior (dark line) and logarithmic prior (light line).}
\end{figure}

We define a set of effective transverse velocities $\{V_{eff}\}$. 5000 simulations of the monitoring data were produced at each effective transverse velocity. An average histogram is also computed from each set of mock observations. For each mock observation at each pre-defined effective transverse velocity $V_{eff}$, a mock measurement of effective transverse velocity $v_{eff}$ is made by minimising the KS difference between the average histogram for $v_{eff}$ and the histogram of the mock observation. Thus we calculate the function representing the likelihood $p_{lhood}(v_{eff}|V_{eff})$ for observing $v_{eff}$ given an assumption for the true value ($V_{eff}$). Similarly, we find the effective transverse velocity ($v_{obs}$) that best describes the observed microlensing rate. Using Bayes' theorem we calculate the posterior probability that the effective galactic transverse velocity $V_{gal}$ is less than an assumed value $V_{eff}$:

\begin{equation}
P_{v}(V_{gal}<V_{eff}|v_{obs})=\int_{0}^{V_{eff}}p_{lhood}(v_{obs}|V_{eff}')p_{prior}(V_{eff}')\,dV_{eff}'.
\end{equation} 
Our results are consistent for the assumptions of flat ($p_{prior}(V_{eff})\propto dV_{eff}$) and logarithmic ($p_{prior}(V_{eff})\propto\frac{dV_{eff}}{V_{eff}}$) prior probabilities for $V_{eff}$. $P_{v}(v_{gal}<V_{eff}|v_{obs})$ is shown in the left hand plot of figure \ref{velmass}. The model presented here and in section \ref{mass_lim} assumes a trajectory parallel to the C-D axis and no smooth matter. $V_{eff}$ is limited to be less than $1000\sqrt{\langle m\rangle}\,km\,sec^{-1}$ with 95\% confidence. 

\section{Determination of lower limit to microlens mass}
\label{mass_lim}

The velocity dispersion in the centre of the lensing galaxy is $\sigma_*\sim 215\,km\,sec^{-1}$ (Foltz et al. 1992). This internal motion produces a microlensing rate $\propto \langle m\rangle^{-\frac{1}{2}}$, thus there is a minimum $\langle m\rangle$ which is consistent with observations.

\subsection{Microlensing due to random proper motions}
We define the equivalent transverse velocity $V_{equiv}$ as the effective transverse velocity of a static screen of point masses which produces a microlensing rate closest to that of the same model with $V_{equiv}$ replaced by the combination of a random velocity dispersion ($\sigma_*$) plus a transverse velocity ($V_{tran}$) (Wyithe, Webster \& Turner 1999a,b). The microlensing rates are considered equivalent when the histograms are most similar (have the minimum KS difference). Note that this process can be inverted to find the $V_{tran}$ corresponding to a measured limit on $V_{eff}$.

\subsection{Limits on the microlens mass}

$V_{equiv}$ quantifies the minimum level of microlensing that must be observed for an assumed physical transverse velocity of the lensing galaxy. Since the measured effective transverse velocity scales with $\sqrt{\langle m\rangle}$ the cumulative probability for $\langle m\rangle$ is 
\begin{equation}
U\left(\langle m\rangle<M=\left(\frac{V_{equiv}}{V_{eff}}\right)^{2}\right)=\int \left(1-P\left(V_{gal}<V_{eff}\sqrt{M}\right)\right)\,p_{prior}\left(V_{eff}\right)dV_{tran}
\end{equation}
(Wyithe, Webster \& Turner 1999c), where the priors $p_{prior}(V_{eff})$ are those used for the computation of $P_v(V_{gal}<V_{eff}\sqrt{M})$.
Note that this distribution (shown in the right plot of figure \ref{velmass} for the two assumed priors) does not represent the mass function. $\langle m\rangle$ is greater than $0.1M_{\odot}$ (95\% confidence), and has a most likely value consistent with that of low-mass stars.

\acknowledgments I would like to thank Rachel Webster and Ed Turner for their input and guidance.

\end{document}